\documentclass[draft]{agujournal2019}
\usepackage{url} %this package should fix any errors with URLs in refs.
\usepackage{acronym}
\usepackage{amssymb}

\usepackage{soul}

\usepackage{siunitx}
\usepackage{lmodern}

\draftfalse
\journalname{JGR: Space Physics}
\begin{document}

\begin{acronym}[printonlyused, nolist]
    \acro{FAC}{field-aligned current}%
    \acro{DifA}{Diffuse aurora}%
    \acro{ZI}{Zone I}%
    \acro{ZII}{Zone II}%
    \acro{MI}{magnetosphere-ionosphere}
    \acro{JEDI}{Jupiter Energetic-particle Detector Instrument}%
    \acro{JADE}{Jovian Auroral Distributions Experiment}%
    \acro{UVS}{Juno Ultraviolet Spectrograph}%
    \acro{MAG}{Juno Magnetometer}%
    \acro{FGM}{Fluxgate Magnetometer}%
    \acro{UV}{Ultraviolet}%
    \acro{JRM09}{Juno Reference Model through Perijove 9}%
    \acro{JRM33}{Juno Reference Model through Perijove 33}%
    \acro{SysIII}{Jupiter System III coordinates}%
    %\acro{Con81}{Connerney 1981 current sheet model}%
    \acro{Con2020}{Connerney 2020 current sheet model}%
    \acro{PSD}{Power Spectral Density}%
    \acrodefplural{PSD}{Power Spectral Densities}
    \acro{rms}{root-mean-square}%
\end{acronym}
\newpage
%%%%%%%%%%%%%%%%%%%%%%%%%%%%%%%%%%%%%%%%%%%%%%%
%% Title:
\title{Investigating Magnetic Field Fluctuations in Jovian Auroral Electron Beams}

%%%%%%%%%%%%%%%%%%%%%%%%%%%%%%%%%%%%%%%%%%%%%%%
%% Authors:
\authors{A. Salveter\affil{1}, J. Saur\affil{1}, G. Clark\affil{2}, A. Sulaiman\affil{3}, B. H. Mauk \affil{2}, J. E. P. Connerney \affil{4}, B. Bonfond \affil{5}}
\affiliation{1}{Institute of Geophysics and Meteorology, University of Cologne, Cologne, Germany}
\affiliation{2}{The Johns Hopkins University Applied Physics Laboratory, Laurel, Maryland, USA}
\affiliation{3}{School of Physics and atronomy, Minnesota Institute for Astrophysics, University of Minnesota, Minneapolis, MN, USA}
\affiliation{4}{Space Research Corporation, Annapolis, USA}
\affiliation{5}{STAR Institute, Université de Liège, Belgium}

%%%%%%%%%%%%%%%%%%%%%%%%%%%%%%%%%%%%%%%%%%%%%%%
%% Corresponding Author:
\correspondingauthor{Annika Salveter}{annika@salveter.net}

%%%%%%%%%%%%%%%%%%%%%%%%%%%%%%%%%%%%%%%%%%%%%%%
%% keypoints:
\begin{keypoints}
\item Large magnetic field perturbations coincide with \ac{UVS} intense auroral arcs
\item Magnetic field fluctuations on temporal scales $<2$~s are only resolvable above $4$~R$_J$
\item Magnetic field fluctuations at $<2$~s above $4$~R$_J$ coincide with auroral main emission zone
\end{keypoints}

%%%%%%%%%%%%%%%%%%%%%%%%%%%%%%%%%%%%%%%%%%%%%%%
%% abstract:
\begin{abstract}

The Juno spacecraft provides a unique opportunity to explore the mechanisms generating Jupiter's aurorae. Past analyses of Juno data immensely advanced our understanding of its auroral acceleration processes, however, few studies utilized multiple instruments on Juno in a joint systematic analysis. This study uses measurements from the Juno Ultraviolet Spectrograph (UVS), the Jupiter Energetic particle Detector Instrument (JEDI), and the Juno Magnetometer (MAG) from the first 20 perijoves. On magnetic field lines associated with the diffuse aurora, we consistently find small-scale magnetic field fluctuations with amplitudes of up to \SI{100}{nT} on time scales of seconds to \SI{1}{minute}. On magnetic field lines directly linked to the main emission, the electron distribution is field-aligned, mostly broad-band in energy, and accompanied by large-scale magnetic field perturbations of several \SI{100}{nT} on time scales of tens of min (except one case). These large-scale perturbations are generally associated with quasistatic field-aligned electric currents.  Small-scale magnetic fields are not resolved over the main emission zone closer than radial distances 4 Jovian radii due to the digitization limit of the magnetometer. However, in all cases where Juno crosses the main auroral field lines beyond \SI{4}{R_J}, the digitization limit is significantly reduced and we detect small-scale magnetic field fluctuations of \SIrange{2}{10}{nT} consistent with a turbulent spectrum.  Associated energy fluxes projected to Jupiter can exceed 1000 mW/m2. The general broad-band nature of the electron distributions and the consistent presence of small-scale magnetic field fluctuations over the main emission support that wave-particle interaction can dominantely contribute to power Jupiter's auroral processes.

\end{abstract}

%%%%%%%%%%%%%%%%%%%%%%%%%%%%%%%%%%%%%%%%%%%%%%%
\section*{Plain Language Summary}
Jupiter possesses the brightest auroral emission of all planets in the solar system.  One of the key objectives of the Juno mission to Jupiter is to better understand the processes that lead to this bright emission. In our analysis, we simultaneously use data from three instruments on Juno, i.e., from the Ultraviolet Spectrograph, the Jupiter Energetic particle Detector Instrument, and the Juno Magnetometer obtained during the first 20 orbits of Juno around Jupiter.  We find that on magnetic field lines connecting to auroral emission, energetic electrons and magnetic field perturbations associated with electric currents are systematically present. Magnetic field perturbations indicate that both large-scale DC electric currents and small-scale AC currents occur. We find that most auroral electrons are broad-band in nature and often travel towards and away from Jupiter, which together with the consistent presence of AC currents suggests what waves, in particular a type referred to as Alfvén waves, might play a key role in energizing auroral electrons. 

%%%%%%%%%%%%%%%%%%%%%%%%%%%%%%%%%%%%%%%%%%%%%%%
\section{Introduction}
Jupiter possesses the most powerful aurora in the solar system. Understanding the mechanisms that energize the electrons and ions causing auroral emissions is a key objective of the Juno mission \cite{Mauk2013}.  Initial observations based on \ac{JEDI} and \ac{JADE} instruments of the Juno spacecraft have shown that electron energy distributions, ranging from low to high energies (\SI{30}{eV} to \SI{1}{MeV}) are often broad and follow power-law distributions with a hardtail that extends to high energies of around \SI{1}{MeV} and energy fluxes exceeding \SI{3}{W/m^2} \cite{Mauk2017, Mauk2020, Allegrini2017, Szalay2017}. In a statistical analysis, the occurrence of these broad distributions was further investigated, revealing that they are present approximately $93(\pm 3.8)\%$ of the time above the main emission region during the first $20$ perijoves \cite{Salveter2022}. In contrast, monoenergetic distributions, where intensities peak at specific energy levels, are rare and account for approximately $7$ $(\pm 3.8)\%$ of the observed times \cite{Salveter2022, Clark2017, Mauk2018, Mauk2022}. Another remarkable observation was the bidirectional nature of electron distributions, indicating electrons accelerated in both directions along magnetic field lines. This phenomenon is particularly prominent in the \ac{ZII} region, identified by \citeA{Mauk2020}, characterized by dominant upward and bidirectional electron distributions associated with a downward electric current region. In contrast, equatorward \ac{ZI} exhibits mostly downward-going electrons linked with upward currents and transitions further equatorward into diffuse auroral regions \acs{DifA} with mostly pancake distributions. Poleward of the main emission in the polar region, a large and dynamic region, energetic electrons move primarily upward along magnetic field lines \cite{Ebert2017, Elliott2018, Mauk2020, Paranicas2018}. Distinguishing the transition from the main emission zone to the polar region proves challenging, as both inverted V and broadband distributions are frequent. 

Jupiter exhibits a unique \ac{MI} interaction with large angular momentum and energy transport between its ionosphere and the plasma of the middle magnetosphere due to the internal mass loading from Io, which is generally thought to ultimately cause the intense aurora.  This contrasts with the aurora on Earth, which is powered by the solar wind \cite{Akasofu1981, Baker1996}. Jupiter’s intense auroral emissions reach an ultraviolet brightness of several thousand kilorayleighs (kR) \cite{Grodent2003} that exceeds the terrestrial auroral brightness by a factor
of $10$ \cite{Broadfoot1979}. At Earth, the dominating auroral type in contrast to Jupiter is diffusive aurora with a smaller contribution from discrete aurora. At Earth, the latter is usually caused by a larger fraction of mostly downward mono-energetic electrons compared to broad-band electron distributions \cite{Newell2009}. Another difference lies in Jupiter's downward current region \ac{ZII}, which shows strong auroral emissions, unlike Earth, where such emissions are absent \cite{Mauk2020}.

Jupiter’s powerful magnetic field, the rapid rotation of the planet, and Io’s plasma production result in a corotating plasma sheet confined to the equatorial region. However, the plasma's corotation cannot be sustained in the middle magnetosphere, leading to magnetic field tension and Lorentz forces, enforcing corotation in the equatorial sheet. The forces are balanced in Jupiter's ionosphere through collisions with the neutral atmosphere. The associated transport of angular momentum and energy from the ionosphere into Jupiter's magnetosphere is associated with a quasi-static electric current system along Jupiter's field lines \cite{Hill1979, Kivelson2005}. This large-scale electric current system has traditionally been associated with the main auroral emission \cite{Hill2001, Cowley2001}. On field lines where the upward electric current maximizes, i.e., around \SI{30}{R_J}, electrons are expected to accelerate towards Jupiter. Using the Knight relationship \cite{Knight1973} for Jupiter, \citeA{Cowley2001} predicted mono-energetic, unidirectional auroral electrons with energies around \SI{100}{keV}. The relationship between electrostatic potential, field-aligned currents, and the resulting auroral excitation has been further studied, e.g., by \citeA{Nichols2004, Ray2010}.

Jupiter's equatorial magnetosphere also exhibits small-scale magnetic field fluctuations on the order of several \si{nT}, which show turbulent properties \cite{Saur2002, Saur2004, Tao2015, Ng2018, Lorch2022, Kaminker2024}. These small-scale magnetic field fluctuations have been associated with Alfvén wave packets and highly time-variable field-aligned electric currents whose dissipation at high latitudes was suggested to contribute to auroral acceleration \cite{Saur2003}. \citeA{Thomas2004, Mauk2007} observed in the middle magnetosphere of Jupiter broadband highly field-aligned electron beams, which were argued to be related to anti-planetward auroral acceleration due to small-scale electric current systems \cite{Mauk2007} similar to the findings and discussion about Saturn's magnetospheric electron beams \cite{Saur2006, Masters2022}. The highly time-variable magnetic fields and associated DC electric currents are related to plasma waves such as kinetic/inertial Alfvén waves. When these waves encounter low plasma densities in the auroral acceleration region such that the scales of the plasma wave approach typical kinetic plasma scales, i.e. the electron inertial length scales \cite{Saur2018, Sulaiman2022}, then wave-particle interaction can take place \cite<e.g.>{Lysak1996}. In addition to (turbulent) Alfvénic fluctuations \cite{Saur2018, Damiano2019} other mechanisms are also expected to lead to a strong wave-particle interaction, such as those originating from field line resonances \cite{Lysak2020} or the ionospheric Alfvén resonator \cite{Lysak2021}.

Although there are no direct electric field measurements to constrain auroral acceleration mechanisms \cite<except for high-frequency plasma waves by the Waves instrument;>{Kurth2017}, Juno’s magnetic field data can suggest the presence of various types of related electric current and possibly associated electric fields. \citeA{Kotsiaros2019} analyzed perpendicular magnetic field fluctuations at high latitudes, providing evidence for quasi-steady state Birkeland currents. Such a large-scale current system might be associated with large potentials \cite{Kotsiaros2019}. The derived currents in \citeA{Kotsiaros2019} are weaker than anticipated and were argued to not fully account for the expected currents related to Jupiter's aurorae, despite their amplitude matching the radial currents detected within the equatorial plane. \citeA{Bonfond2020} provided multiple arguments challenging the notion that Jupiter's main aurora is primarily caused by the large-scale corotation enforcement current.  For example, the aurora appears to be more intense at dusk, whereas the radial currents are more pronounced at dawn \cite{Groulard2024}. In contrast, small-scale magnetic field fluctuations and time-variable electric currents indicating wave-particle interactions have been observed at high latitudes as well \cite{Gershman2019, Sulaiman2022}. \citeA{Gershman2019}, e.g., observed that small-scale perpendicular magnetic field fluctuations occurred along with auroral broadband emissions and the Poynting flux reaching up to \SI{100}{mW/m^2}. Small-scale magnetic field fluctuations have previously been detected primarily over diffuse auroral regions but appeared to be not present as the spacecraft passes through the main emission zone \cite{Sulaiman2022}. The authors suggested that the diminished magnetic fields resulting from very small plasma densities, known as auroral cavities, might explain the lack of fluctuations over \ac{ZI}. This density drop leads to large electron inertial length scales, increasing the wave-particle interaction of kinetic/inertial Alfvén waves thereby reducing the magnetic field fluctuations while accelerating auroral particles \cite<e.g.>{Lysak2021, Saur2018}. However, evidence for Alfvénic turbulence at mid to high latitudes, characterized by small-scale magnetic field fluctuations and sufficient Poynting fluxes to drive auroral emissions, has been found at radial distances of \SI{>10}{R_J} by \citeA{Lorch2022}. 

To further shed light on the dominant mechanisms that cause Jupiter's auroral electrons associated with the diffuse and main emission, this work presents a statistical study in which we jointly analyze auroral electron intensity from \acf{JEDI}, and magnetic field variations from \acf{MAG}, and compares these with ultraviolet emissions observed by \acf{UVS}. By examining magnetic field changes, we investigate also the role of the possible electric currents associated with the observed particle distribution and auroral patterns. In Section \ref{sec:data}, we briefly present the important properties of the \ac{JEDI}, \ac{JADE}, and \ac{MAG} instruments, and the data and analysis tools applied in this study. In Section \ref{sec:results} we present the results of our data analysis, categorized by radial distance from Jupiter: within \SI{4}{R_J} and at greater distances. At greater distances, digitization of the magnetometer significantly enhances the resolution of small-scale magnetic field fluctuations. In Section \ref{sec:summary}, we end with a summary of our main conclusions. 

%%%%%%%%%%%%%%%%%%%%%%%%%%%%%%%%%%%%%%%%%%%%%%%
\section{Instrument, Data and Methods}\label{sec:data}
To gain a deeper understanding of the distinctive features of Jupiter's magnetosphere at high latitudes, particularly in relation to the auroral regions, we have integrated data from three instruments aboard the Juno spacecraft. During the first 20 perijoves, the spacecraft has conducted in situ measurements while passing Jupiter at low altitudes and high latitudes above the Jupiter auroral region. These unique data enable us to gain a broader understanding by statistically analyzing observations from multiple perijoves and various magnetospheric properties, rather than focusing solely on individual observations. Now, we briefly introduce the three instruments used in this study—the \acf{JEDI}, \acf{UVS}, and \acf{MAG}—along with the basic processing routines we employed.

    \subsection{Jupiter Energetic Particle Detector (JEDI)}\label{subsec:JEDI}
The \ac{JEDI} instrument, in conjunction with its low-energy counterpart \ac{JADE}, carries out in situ observations of electron and ion distributions \cite{Bagenal2017}. \ac{JEDI} measures the precipitation of energetic electrons within the energy range of $25$ to $1200$~keV \cite{Mauk2017}. It is equipped with 18 solid state detectors (SSD) that simultaneously measure the rate of single electrons in different directions, providing an almost complete \ang{360} field of view \cite{Mauk2017}. This allows for the measurement of all pitch angles, which is the angle between the velocity of the electrons and the magnetic field vector, and leads to a nearly full pitch angle coverage. The measurements are obtained every \SI{0.5}{s}, enabling the resolution of even small structures of a few hundred kilometers because of the spacecraft's velocity of approximately $50$~km/s near Jupiter. However, the pitch angle coverage of \ac{JEDI} depends on the alignment of the magnetic field with the plane perpendicular to the spacecraft spin vector, which means that reasonable pitch angle coverage is sometimes only achieved over the $30$~s spacecraft spin period. Given a complete population of the loss cone, the energy flux can be calculated and projected onto the atmosphere by adding $\pi \sum_{n} I_n(\alpha) E_n \Delta E_n$, where $\pi$ represents the area projection-weighted size, $n$ refers to individual energy channels with $E_n$ as central energy, $\Delta E_n$ as the energy bandpass and $I_n$ the measured intensity \cite{Mauk2017b}.

    \subsection{Juno Ultraviolet Spectrograph (UVS)}\label{subsec:UVS}
\begin{figure}[htb]
    \centering
    \includegraphics[width=\textwidth]{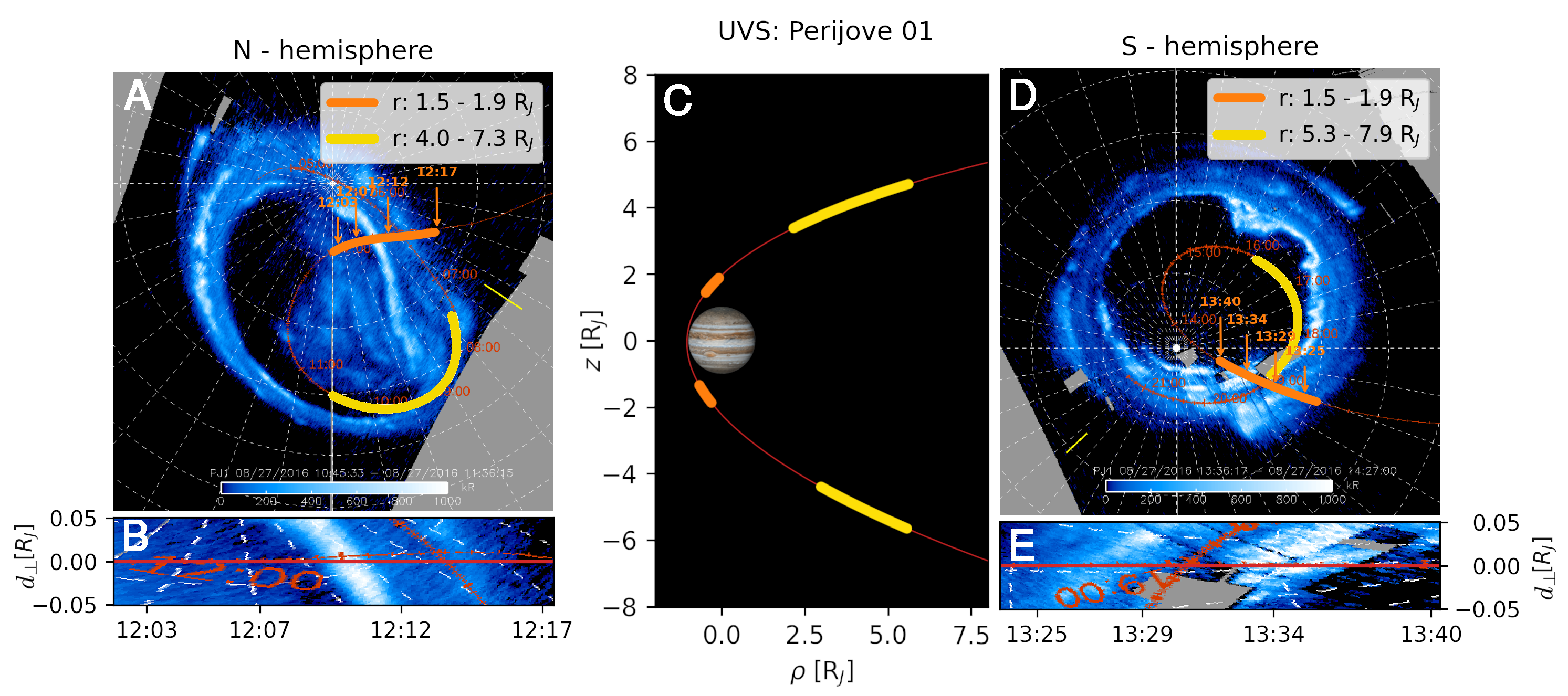}
    \caption[Main emission crossing exemplary shown for Perijove 1]{Overview of the flight trajectory in panel C of the first perijove in the rotating frame of Jupiter, expressed by the \ac{SysIII} coordinates. The perpendicular distance from the spin axis is calculated by $\rho=\sqrt{x^2+y^2}$. The spacecraft's location is mapped onto the ionosphere and marked with red lines on the left and right sides of the \ac{UVS} panels. The \ac{UVS} images are shown for the northern hemisphere (A) on the left and the southern hemisphere on the right side (D). The lower parts (B and E) show the \ac{UVS} image slice along the flight trajectory for both hemispheres, where Juno crosses the expected auroral emission region at times indicated by the orange and yellow lines in the top plots, at low and high altitudes, respectively. Some regions may lie above the main emission region, but those beyond \SI{8}{R_J} radial distance are excluded due to the resolution limits of the instrument. The perpendicular distance to the flight trajectory is given by $d_{\perp}$ in Jovian radii.}
    \label{fig:data_coverage}
\end{figure}

The Ultraviolet Spectrograph (UVS) observes ultraviolet emissions from Jupiter, covering wavelengths from \SI{68}{nm} to \SI{210}{nm} \cite{Gladstone2017}. \citeA{Bonfond2021} introduces a method for comprehensive imaging of the whole northern and southern aurora in polar projection, at an altitude of \SI{400}{km} above the planet's 1-bar level, where the detector distinguishes ultraviolet photons from background interference. During approximately $50$ minutes, the successive rotations update the \ac{UV} image by calculating the weighted average, prioritizing recent ones, resulting in a comprehensive overview of the whole polar-projected auroral region, demonstrated in panels A and D of the first perijove in Figure \ref{fig:data_coverage}.
To associate particular \ac{UV} emission patterns with the spacecraft's location along magnetic field lines, a mapping technique created by \citeA{Wilson2023} is used. This method utilizes both the internal \ac{JRM33} \cite{Connerney2022} and the external magnetic field models \ac{Con2020} \cite{Connerney2020}. 
A set of lines parallel to Juno’s trajectory is determined to encompass a curved section along the trajectory of the spacecraft. This curved section is extracted from the averaged polar \ac{UVS} images to create a two-dimensional rectangular representation of the averaged polar \ac{UVS} along the line of flight (see panels B and E in Figure \ref{fig:data_coverage}). The effectiveness of this method's display has been evaluated using the standard Phase Alternating Line (PAL) test pattern \textit{Philips PM5544}. This pattern comprises elements such as color bars, grayscale, and geometric shapes, which are utilized to assess parameters like color fidelity, contrast, sharpness, alignment, and aspect ratio. Achieving a satisfactory representation requires at least $40$ lines per \SI{0.1}{R_J} for optimal data visualization.
% LIMITATIONS of slicing

Discrepancies between the location of the time-averaged observation of the ultraviolet emissions and the instantaneous in-situ measurements taken by \ac{MAG} and \ac{JEDI} may arise due to various factors. First, the $50$ minute duration for capturing average polar \ac{UVS} images introduces timing uncertainties for time-variable auroral \ac{UV} emissions. Second, changes in the magnetic field over time may affect the accuracy of tracing field lines. Third, the curved flight path can distort \ac{UV} emission structures when projected onto a rectangular shape, as is exemplarily shown in panels \ref{fig:data_coverage} B and \ref{fig:data_coverage}E.  Nevertheless, the resulting \ac{UVS} image slices offer a comprehensive view of \ac{UV} emission morphology for comparison with magnetic field configurations and electron intensity.
    
\subsection{Fluxgate Magnetometer}
To observe magnetic field fluctuations along the flight trajectory, data from the \acf{FGM} is utilized. Operating at 64 samples per second, the \ac{FGM} measures magnetic field magnitude and direction, enabling observation of variations on short temporal and spatial scales in all three dimensions simultaneously \cite{Connerney2017a}. Additionally, an onboard star camera observes the position of fixed stars to obtain the spacecraft orientation, achieving an accuracy of \SI{100}{ppm} for magnetic field vectors.

The magnetometer is capable of sensing magnetic field strengths within a range of a few nanoteslas (\si{nT}) up to $16$ Gauss (equivalent to \SI{1.6e6}{nT}) per axis, operating in one of $6$ dynamic ranges in response to the measured field magnitude \cite{Connerney2017a}.
The analog output in each range is sampled by a 16-bit analog-to-digital converter (A/D), resulting in a resolution of \SI{0.025}{nT} in the instrument’s most sensitive dynamic range (\SI{\pm 1600}{nT}), increasing (by a factor of $4$) as the dynamic range increases by a factor of $4$, to a resolution of \SI{25}{nT} in the $16$ Gauss dynamic range. The \acf{FGM} possesses a natural noise level of less than \SI{\ll 1}{nT}, which is considered negligible compared to the uncertainty due to digitization. 

The calibrated magnetic field data for this study was obtained from the Planetary Data System \cite{System2022} in planetocentric coordinates. The data is provided with a temporal resolution of $64$ samples per second near perijove, dropping to $16$ samples per second further from Jupiter and ultimately $8$ samples per second throughout the remainder of Juno’s highly elliptical orbit, to meet the telemetry allocation constraints.

\subsection{Magnetic Field Fluctuations}\label{subsec:limits}
Fluctuations in the magnetic field caused by effects of temporal or spatial variability are studied by looking closely at the residual that remains after subtraction of the model fields, typically no more than about \SIrange{0.1}{1}{\%} of the ambient field. 
Thus, the initial data processing step involves subtracting the background magnetic field defined by the internal magnetic field model \ac{JRM33} and the external magnetic field model \ac{Con2020}. 

Background magnetic field data are obtained from a module developed by \citeA{Wilson2023}, which offers both internal and external magnetic field components in \ac{SysIII} coordinates based on \ac{JRM33} and \ac{Con2020}. The magnetic field variations are then given in the Jupiter coordinate system \ac{SysIII} with $\delta B_i$ along the longitudinal $\phi$, latitudinal $\theta$, and radial $r$ axes. 

%%%%%%%%%%%%%%

The calibrated magnetometer data in the high dynamic ranges used in this study (ranges $4,5,6$) have been corrected for a spacecraft field arising from Eddy currents due to the spacecraft rotation in a strong magnetic field \cite{Kotsiaros2020}. The corrections applied are described in Addendum $1$ to the Software Interface Specification \cite{Connerney2017a} and are a few hundred parts per million in magnitude. However, the residuals for some perijoves contain a recognizable sinusoidal component with a period of \SI{\sim 30}{seconds}, corresponding to the Juno spin period. Spin modulation can arise from a variety of sources, including induced and remanent spacecraft magnetic fields, instrument measurement errors, coordinate transformation errors, and even data quantization uncertainties. However, we make here no attempt to correct these to maintain data integrity; instead, these artifacts are carefully observed during interpretation and do not impact our results.

Changes in the magnetic field are represented by the longitudinal component $\delta B_{\phi}$, the latitudinal component $\delta B_{\theta}$, and the radial component $\delta B_{r}$. In a dipolar-dominated magnetic field at low altitudes and high latitudes, the radial component aligns roughly with the magnetic field direction, while the longitudinal and latitudinal components are roughly perpendicular to a dipole-dominated field. Thus, variations in the $\theta$ and $\phi$ components indicate magnetic field perturbations perpendicular to the background field, often associated with parallel electric currents. In the following, we will analyze magnetic field fluctuations in a broad frequency range up to \SI{2}{Hz}.

\subsection{Field-aligned Static Currents}\label{subsec:static}
%MAGNETIC FIELD CHANGES direction
To examine the contribution of magnetic field fluctuations that indicate field-aligned current density from radial currents in the ionosphere, we apply Ampère's law, $\nabla \times \vec{B} = \mu_0 \vec{j}$, where $\mu_0$ is the vacuum permeability while neglecting displacement currents. The resulting magnetic field changes are then associated with the field-aligned currents by
\begin{equation} 
    j_{\parallel} \approx \pm j_r = \frac{\pm 1}{\mu_0 r \sin \theta}\left(\frac{\partial (\delta B_{\phi}\sin\theta)}{\partial\theta}-\frac{\partial (\delta B_{\theta})}{\partial\phi}\right), \label{eq:currentdensity}
\end{equation}
where $\delta B_{\phi}$ corresponds to the background subtracted latitudinal magnetic field and the $\pm$ signs to the northern and southern hemispheres, respectively.
We assume in the following that the current density $j_r$ at high latitudes and low altitudes represents the field-aligned currents.
The measurements obtained by the Juno spacecraft do not permit independent determination of the $\theta$ and $\phi$ derivatives in equation (\ref{eq:currentdensity}). However, assuming that the electric currents connecting Jupiter's magnetosphere with its ionosphere are structured in sheets with locally very small variability in the $\phi$-direction, but only in the $\theta$- direction (such as the \ac{MI}-current system suggested by \citeA{Hill2001, Cowley2001}), then we can neglect to first order $\frac{\partial (\delta B_{\theta})}{\partial \phi}$. This approximation is expected to hold for the large-scale \ac{MI}-electric current system \cite{Hill2001, Cowley2001}.
It is important to note that single spacecraft observations are limited to measuring changes only along the spacecraft's path. Therefore, if the current sheet is not perfectly perpendicular to the trajectory of the spacecraft, the variations perpendicular to it could be underestimated, affecting the resulting field-aligned currents \cite{Luhr1996}. Note, we do not apply \ref{eq:currentdensity} in our study, but only discuss this equation to help with the interpretation of observed large-scale magnetic field perturbations.

The polarity, i.e. direction, of the magnetic field change is directly linked to the direction of the field-aligned current, which can either be directed towards or away from Jupiter. 
For measurements as a function of time in a moving spacecraft, the polarity depends on the direction of the trajectory, whether it is towards the poles or the equator, and the orientation of the magnetic field, which is upward in the northern hemisphere and downward in the southern hemisphere. Analyzing the magnetic field components in the \ac{SysIII} coordinates, we find that negative magnetic field changes indicate upward currents when Juno is heading toward the poles in the northern hemisphere. Each alteration in the flight path or observation hemisphere will lead to a reversal of the sign to determine the direction of the field-aligned electric current and must be taken into account.

Substantial changes in the magnetic field with considerable amplitudes over several minutes are likely linked with \ac{FAC} and will be termed \textit{ large-scale magnetic field variations}. These variations are characterized by isolated structures associated with periods greater than one minute, corresponding to frequencies less than \SI{0.01}{Hz}. Thus, Juno detects these variations in the magnetic field over distances exceeding \SI{3000}{km}. Conversely, the high-frequency magnetic field changes with periods shorter than seconds, and hence frequencies above \SI{0.1}{Hz} are termed \textit{small-scale fluctuations} henceforth. Their amplitudes are smaller and sometimes indistinguishable from background noise. However, these small-scale fluctuations can occur over an extended time interval of several minutes. The subsequent section will explore the potential physical significance and occurrence of these small-scale fluctuations.

\subsection{Field-aligned Non-Static Currents}
%%%%%%%%%%%%%%%%%%%%%%%%%%%
To explore the fluctuations in the magnetic field at high frequencies, potentially arising from dynamic \ac{MI}-coupling and associated time-variable field-aligned currents, it is important to distinguish between high-frequency reaching \SI{1}{Hz} and low-frequency fluctuations with less than \SI{1}{mHz}, as well as to filter out noise by focusing on specific frequencies of interest. 
To study the temporal evolution of the frequency components of a signal with $N$ samples and the sampling interval $\Delta t$, we use the wavelet analysis which convolves the signal with localized basis functions in both the time and frequency domains. The resulting Continuous Wavelet Transform (CWT) coefficients $W_n(s)$, conserve the average signal energy during the transition from the time to the frequency domain, following Parseval's relation \cite{Torrence1998, Paschmann1998}. To ensure energy consistency across interval modifications, the \acf{PSD} is given by $S_n = 2\frac{N}{f_s} \cdot W_n(s) ^2$ in units of \si{(nT)^2/Hz}. A Morlet basis function $\Psi_0(\eta) = \pi^{-1/4}\exp{i\omega_0\eta}\exp{-\eta^2/2}$ is used based on sinusoidal waves with a Gaussian envelope, with a nondimensional "time" parameter $\eta$ and the nondimensional frequency $\omega_0$ representing the number of oscillations in a wave packet. Within these studies, we choose the nondimensional frequency $\omega_0=6$, which establishes an almost direct relationship between the wavelet scale and the Fourier period $\lambda=1.03 s$ \cite<refer to Table 2 for other empirical values in >{Torrence1998}.

To remove any unresolved low-frequency trends/noise in the data, the data undergo pre-whitening before the wavelet transform and post-darkening afterward. Pre-whitening involves computing the time increment series of the data as $b(t, \tau) = B(t + \tau) - B(t)$, where $\tau$ represents the time increment of the series. The post-darkening step guarantees energy conservation by multiplying the Power Spectral Density (\ac{PSD}) by the factor $(4\pi \sin^2(\pi f\Delta t))^{-1}$ \cite{Bieber1993}. To reduce edge effects in the wavelet transform, zeros are added to the signal until the next power of two is reached. 
Yet, edge effects are visible in the Cone-of-Influence (COI), which refers to the region where the sidelobes of the Morlet wavelet exceed $e^{-1}$ when reaching the data boundary \cite<for detailed information>{Torrence1998}. Consequently, we extend the time intervals of interest to eliminate the boundary effects on the time periods specified for the frequencies above \SI{1}{mHz}.

\begin{figure}[htb]
    \centering
    \includegraphics[width=0.9\textwidth]{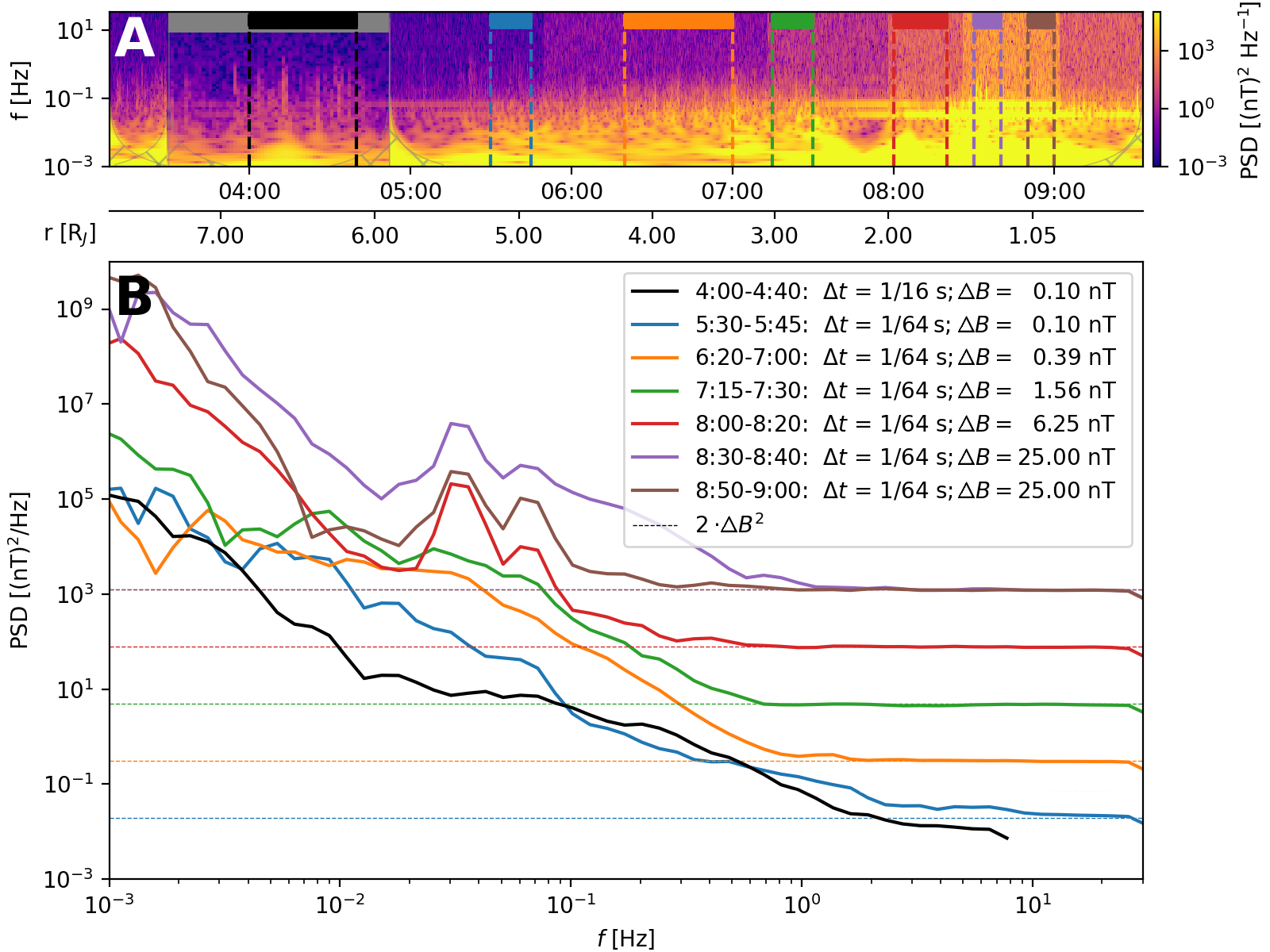}
    \caption[Magnetic field digitization levels for the Perijove 5 flyby]{Wavelet spectrum of the northern flyby of Fifth Perijove are displayed over a wide temporal range of more than six hours in part (A). All the different digitization levels are visible. (B) Time-averaged Power Spectral Densities (PSD) during selected time intervals, with the dashed lines indicating the corresponding digitization levels.}
    \label{fig:PSD_digitization}
\end{figure}

To distinguish regions with significant fluctuations at either small or large scales, the wavelet power is averaged over specific frequency bands, indicating the average variability within a specific frequency range using Parseval's Theorem \cite{Torrence1998}. The \acf{rms} value is derived by taking the square root of $S_n$ the \ac{PSD}. Consequently, the local power spectra for specific time windows are calculated by averaging over the frequency bands of the wavelet spectra, as illustrated in Figure \ref{fig:PSD_digitization} B, with respect to time periods of various digitization levels, indicated in Figure \ref{fig:PSD_digitization} A.

Resolving fluctuations associated with higher frequencies (small-scale) fluctuations is subject to limitations due to the restricted resolution of \ac{FGM} data caused by digitization.

\begin{figure}[htb]
        \centering
        \includegraphics[width=\textwidth]{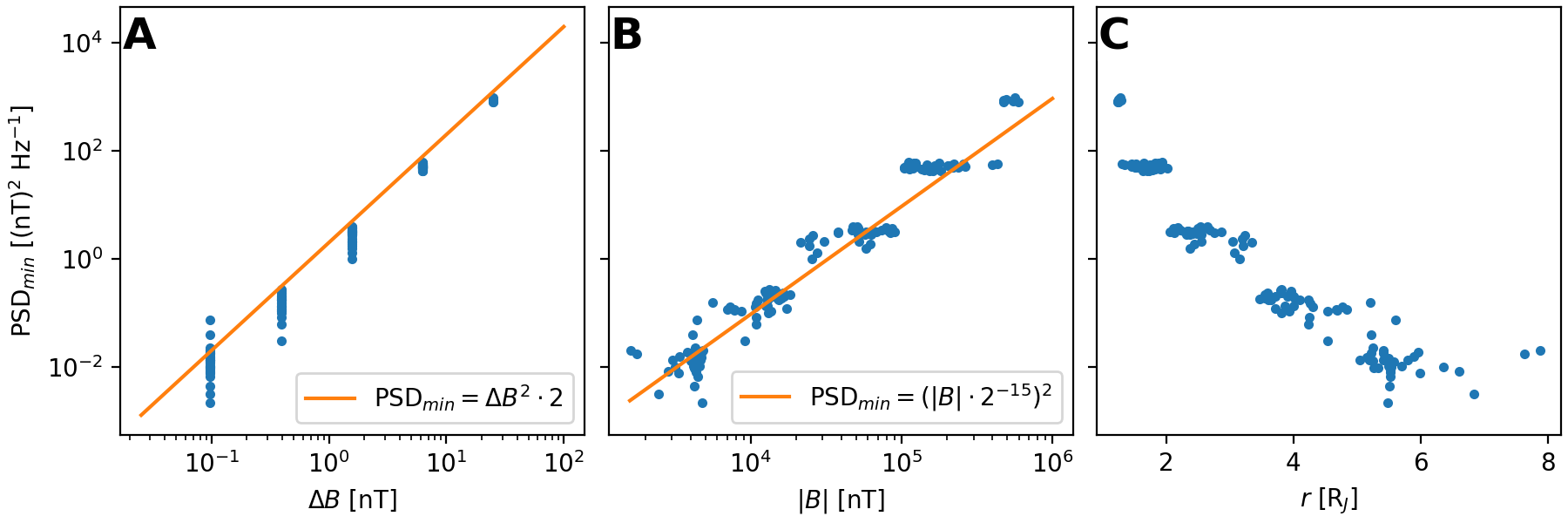}
        \caption{The minimum resolvable \acl{PSD} $PSD_{min}$ is analyzed for each digitization level (panel A), demonstrating its correlation with the total magnetic field strength (panel B) and the radial distance to Jupiter (panel C).  The orange slopes in panels A and B indicate the expected minimum resolvable \acl{PSD} $PSD_{min}$ for the digitization level $\Delta B$ and the total magnetic field strength $|B|$, respectively.
        }
        \label{fig:correlation_dB_CWT}
    \end{figure}

The quantization error, as described by \citeA{Bennett1948} and \citeA{Gray1998}, exhibits little correlation with the real signal, acting as random noise with a nearly uniform white spectrum.
 Consequently, \acp{PSD} are constrained by the square of quantization noise $\Delta B$, following $2\cdot \Delta B^2$, as depicted by dotted lines in Figure \ref{fig:PSD_digitization} B. This figure illustrates the average power spectral densities determined for different digitization levels in various time periods during the northern perijove 5, marked by colored regions in Figure \ref{fig:PSD_digitization} A, showing limits within each digitization level.
For most levels, frequencies above approximately \SI{0.5}{Hz} are mainly affected by quantization noise. However, the smallest digitization level of \SI{0.072}{nT} can detect changes in amplitude down to \SI{0.1}{(nT)^2/Hz}, allowing the observation of frequencies up to around \SI{3}{Hz} on average. The relationship between the quantization error $\Delta B$ and the lowest detectable \acl{PSD} $PSD_{min}$ is shown in Figure \ref{fig:correlation_dB_CWT} A, where the equation $PSD_{min}= \Delta B^2 \cdot 2$ delineates the minimum detectable \acl{PSD}.

The magnetic field strength is associated with the least resolvable \acl{PSD} $PSD_{min}$ through expression $PSD_{min}= ( B \cdot 2^{-15})^2$, and thus is related to the radial distance to Jupiter, as seen in Figure \ref{fig:correlation_dB_CWT} B. Consequently, observations at radial distances less than \SI{4}{R_J} are unlikely to be able to resolve possible magnetic field fluctuations with amplitudes less than \SI{1}{nT}, which are mostly related to wavelengths less than \SI{2}{\s}. Thus, only observations at a greater radial distance than \SI{4}{R_J} can resolve sufficient frequencies up to \SI{3}{Hz}. 
The highest resolution of magnetic field data is achieved with a quantization step of \SI{0.025}{nT} \cite{Connerney2017a}, typically used during flybys at radial distances greater than \SI{\sim 6}{R_J}. Although this is expected to improve the resolution of higher frequencies, we did not use this range since we focused on examining magnetic field perturbations near Jupiter's poles and above the main emission region.

%%%%%%%%%%%%%%%%%%%%%%%%%%%%%%%%%%%%%%%%%%%%%%%
\section{Results}\label{sec:results}
\begin{figure}[ht!]
    \centering
    \includegraphics[width=0.6\textwidth]{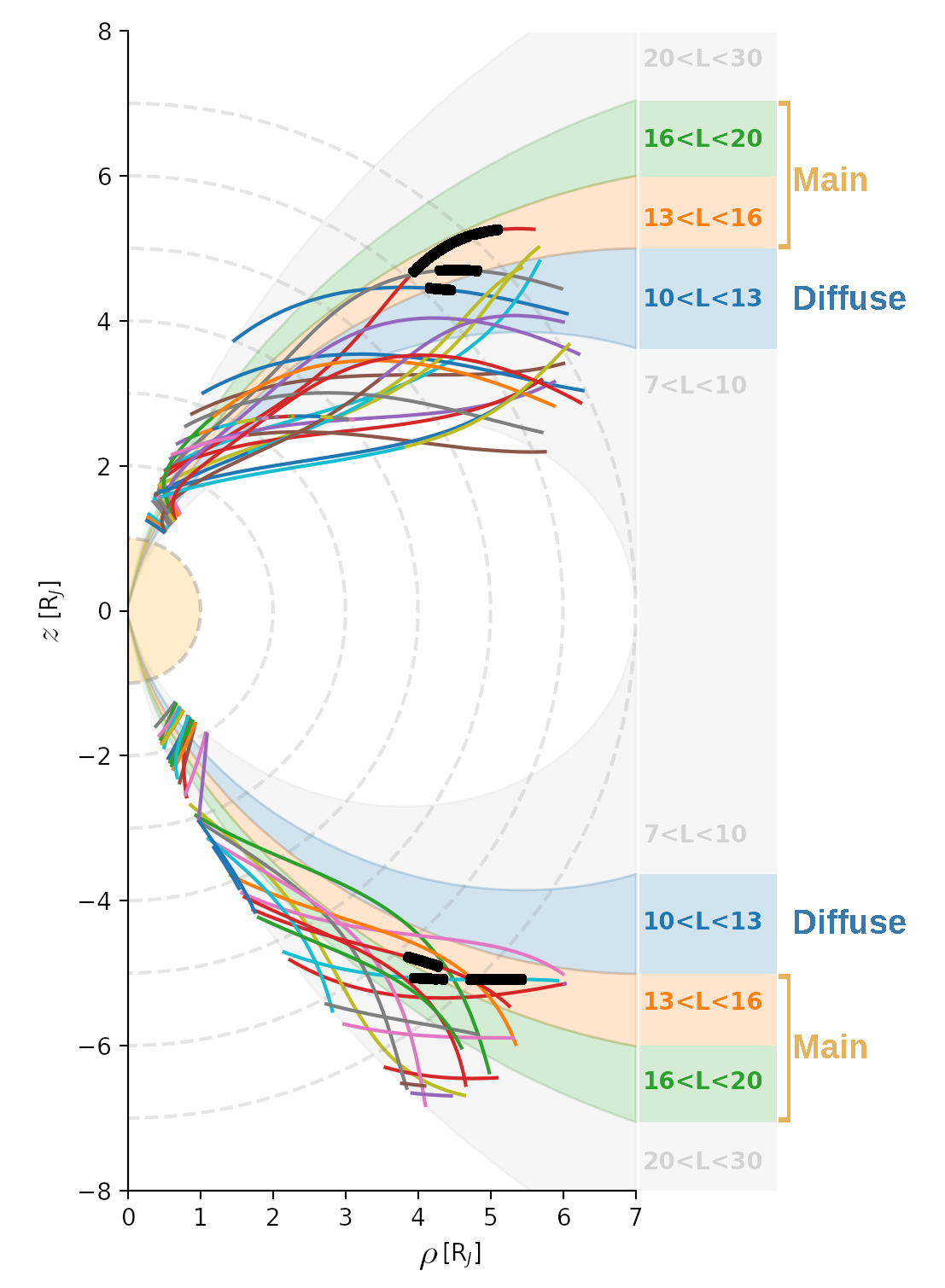}
    \caption{The Figure illustrates the data coverage in \ac{SysIII} coordinates, where the line plots outline the paths observed in the initial $20$ perijoves. $\rho$ represents the distance perpendicular and $z$ represents the distance along the axis aligned with magnetic dipole. The circular dashed lines on the graphs represent radial distances. The green, orange, and blue segments signify regions between the L-shell values of $10$, $13$, and $16$, respectively, indicating the \ac{DifA}, \ac{ZI}, and \ac{ZII} regions. The L-shell approximates the internal magnetic field and neglects the current sheet. It is used as a simple and transparent method to map the magnetic field into Jupiter's atmosphere. Furthermore, black lines indicate observations of small-scale magnetic field fluctuations detected at higher altitudes.}
    \label{fig:data}
\end{figure}

Only about half of all flybys are suitable for a joint analysis of the magnetic fields, energetic electrons, and auroral UV emission at high latitudes. This is because the spacecraft did not cross field lines connected to the main emission — particularly at higher altitudes — or because the field of view is restricted by the spacecraft's rotation. 
In particular, the spacecraft's rotation can cause restrictions on the measurements of the magnetometer and particle instrument. Therefore, some auroral crossings do not provide sufficient data resolution to investigate the correlation between the different instrument observations. 
However, the trajectory of the spacecraft passes through the magnetic field lines mapping to the primary auroral emission multiple times throughout the flybys, as illustrated in Figure \ref{fig:data} by the radial and L-shell coverage of the various flybys. 

To compare the time-averaged \ac{UVS} observations with the instantaneous magnetic field and particle observations, we use the L-shell parameter to localize auroral intensities and acceleration processes.
This parameter was computed using the routine developed by \citeA{Wilson2023}, which maps the magnetic field lines according to the internal model \ac{JRM33} introduced by \cite{Connerney2022}. Since the isolines of the L-shell parameter closely match the auroral shape, it offers a quantitative means of comparing auroral regions across all longitudes. 

For comparison of auroral electron distributions and magnetic field perturbations mapped to specific auroral features, we focus on the L-shell parameter, which approximates the internal field, but neglects Jupiter's current sheet. We use the L-shell for the purpose of associating in-situ observations close to Jupiter with auroral features in Jupiter's upper atmosphere, but not for mapping these observations into equatorial regions of Jupiter's magnetosphere. Previous studies have shown that the L-shell parameter is more robust than the M-shell parameter for our purpose \cite<see>[ and specifically the Supplementary Information of this work, which demonstrates spurious effects that can occur when applying the M-shell parameter]{Salveter2022}.

The closest crossing occurs at a jovicentric radial distance ranging from \SIrange{1.25}{3}{R_J}, while the second closest crossing occurs at a radial distance between \SIrange{1.6}{8.2}{R_J}. The closest approach to the main emission region typically lasts around $15$ to $20$ minutes, whereas the subsequent closest approach extends much longer, averaging $111$ minutes due to the alignment of the spacecraft's trajectory with the magnetic field.
In these encounters, the spacecraft traverses magnetic field lines with L-shell values ranging from \SIrange{8}{20}{} shown in Figure \ref{fig:data}, allowing observation of the three primary emission zones: \ac{DifA}, \ac{ZI}, and \ac{ZII}, indicated by blue, orange, and green areas. 
These regions are roughly indicated using L-shell values, providing a reference for comparing different auroral zones. These definitions are based on observations from the perijoves, where typical L-shell parameters were identified by electron distribution types and UV emission characteristics. However, the L-shell parameter alone is insufficient for precisely delineating the true boundaries of these regions, as they are strongly influenced by spatial and temporal variations in magnetospheric conditions \cite{Head2024}. However, Jupiter's rotation beneath the spacecraft causes more pronounced azimuthal motion along the main emission region at higher altitudes, resulting in fewer perpendicular passes through all three regions, with some flybys remaining either poleward or equatorward of the main emission region without crossing it. Additionally, certain perijoves do not provide sufficient resolution to encompass the entire electron pitch angle space, yet they are still utilized to examine the correlation between \ac{UV} emissions and magnetic field fluctuations.

An important advantage of analyzing magnetic field measurements at higher altitudes is the lower digitization level of \SI{1}{nT}, which is significantly better than the lower resolution of \SI{25}{nT} within radial distances below \SI{3}{R_J}. By examining auroral crossings at radial distances ranging from \SIrange{1.25}{8.2}{R_J}, a thorough qualitative investigation can be conducted to understand the correlations between electron distributions, magnetic field variations, and ultraviolet emissions in the three distinct regions outlined in the following sections.

\subsection{The Polar region} 
\begin{figure}[!ht]
    \centering
    \includegraphics[width=\textwidth]{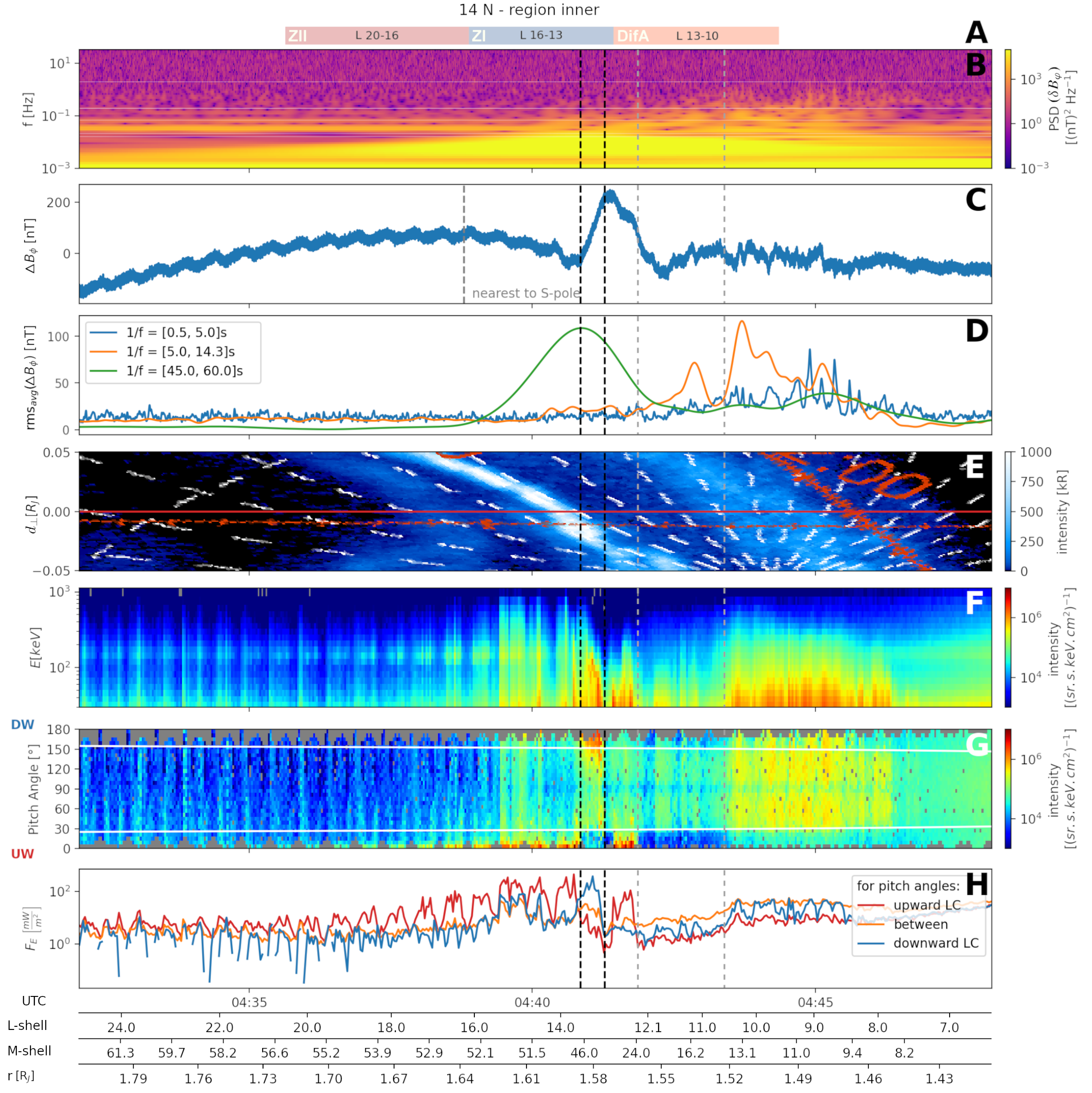}
    \caption[Multi-instrument data for northern Perijove 14 at a radial distance \SI{<2}{R_J}]{Overview of the instrument data from Perijove 14, which crossed the northern hemisphere at altitudes lower than \SI{1}{R_J}. Panel (A) displays the L-shell value color-coded in red, blue, and orange, representing the three anticipated auroral regions: \ac{ZII}, \ac{ZI}, and \ac{DifA} from \cite{Mauk2020}, respectively. (B) Wavelet spectrum of the $\delta B_{\phi}$  component. (C) Azimuthal magnetic field component $\delta B_{\phi}$ obtained by subtracting the \ac{JRM33}+\ac{Con2020} model values from the full magnetic field. Panel (D) shows the \ac{rms} obtained from the averaged \ac{PSD} between the frequency ranges, labeled in the legend. Panel (E) shows the \ac{UVS} observations projected along the Juno flight trajectory. Panels (F) and (G) provide the electron intensity distribution through energy and pitch angles, respectively. The white line in panel (G) corresponds to the loss-cone boundary, which was calculated from the ratio between the magnetic field at the spacecraft and the magnetic footprint in the atmosphere. Lastly, panel (H) provides the mean energy flux for each resulting pitch angle within the loss cones and between. The vertical dashed black lines indicate regions of intense aurora and correspond to strong magnetic field changes. The vertical dashed grey lines indicate regions of empty loss cones and low auroral intensities in the \ac{UV}, but still some magnetic field variations.} 
    \label{fig:Overview_inner_faint}
\end{figure}

The polar region exhibits minimal magnetic field variations at radial distances between \SIrange{1.25}{3}{R_J}, except for some minor noise-related signals, as shown in Figure \ref{fig:Overview_inner_faint} before 4:35 UTC. Furthermore, the high-altitude polar cap crossings between \SIrange{3}{8}{R_J} radial distance, which occur mainly in the southern hemisphere due to the flight trajectory, do not show significant small-scale magnetic field fluctuations, as shown in Supplementary Figure S5. This suggests that any magnetic fluctuations are likely too small to be observed but could still influence acceleration processes through, e.g., whistler-mode waves.

Electron intensities are very low, indicating weak or absent local acceleration, with only a few electrons moving upward or in both directions, or none detected at all. Furthermore, Ultraviolet emissions are faint or absent, suggesting that the spacecraft is in an area with little to no accelerated auroral particles. These observations indicate a quiet magnetospheric environment during the observed periods.

%%%%%%%%%%%%%%%%%%%%%%%%%%%%%%%%%%%%%%%%%%%%%%
%images/02_UVS_1_17_2s_l30keV_16.png

\subsection{The Diffuse Auroral region}
L-shell values smaller than $L < 8$ show no evidence of magnetic-field fluctuations or \ac{UV} emission, and are typically characterized by isotropic electron distributions, as seen in Figure \ref{fig:Overview_inner_faint} after to 4:47 UTC. However,  L-shells between \SIrange{10}{13}{} sometimes reveal faint and indistinct \ac{UVS} emission, lacking a clear structure. The azimuthal magnetic field shows minimal fluctuations, usually comparable to background noise levels. The electron pitch angle distributions typically show pancake distributions characterized by empty loss cones, at times, reduced intensity near \ang{90} resembling a butterfly distribution. The loss cones are typically empty; however, pitch-angle scattering of the trapped particles can lead to partial filling of the loss cones. In such cases, the downward loss cone becomes populated, while the upward loss cone remains relatively empty as those particles are lost to the atmosphere. The energy flux carried by trapped electrons outside of the loss cones typically ranges from \SIrange{10}{100}{mW/m^2} at radial distances of approximately \SIrange{1.3}{1.6}{R_J}, as exemplarily shown in Figure \ref{fig:Overview_inner_faint}. These regions appear to show very limited auroral activity.

On average, at L-shells between approximately $10$ and $13$, accelerated auroral electrons intermittently occupy the downward loss cone, with energy fluxes peaking around \SI{\approx100}{mW/m^2}, both inside and outside the loss cone. This behavior, shown in Figure \ref{fig:Overview_inner_faint}, occurs at lower L-shells between $8$ and $11$ and highlights the challenge of identifying precise boundaries. Nonetheless, the L-shell boundaries are introduced as a rough reference to compare the observations through all perijoves. However, the regions between approximately $10$ and $13$ exhibit a consistent luminosity of about \SI{200}{kR}, along with sporadic patchy \ac{UV} glows, albeit less bright compared to regions associated with higher L-shells. Concurrently, these areas are dominated by small-scale magnetic field fluctuations occurring at intervals between \SIrange{0.5}{60}{\sec} with \ac{rms} values ranging from \SIrange{50}{100}{nT}. The large amplitude magnetic field fluctuations on field lines associated with diffuse aurora are attributed to whistler-mode chorus waves and lead to pitch-angle scattering of trapped electrons \cite{Elliott2021}. Sometimes, an intermediate region with empty loss cones separates the dim \ac{UV} radiations of diffuse aurorae from the intense and luminous auroral emissions originating from \ac{ZI}.

%%%%%%%%%%%%%%%%%%%%%%%%%%%%%%%%%%%%%%%%%%%%%%
\subsection{Strong Electron Beams connected to Auroral Arcs}
Between L-shells \SIrange{13}{16}{}, marked by the blue bar at the top of Figure \ref{fig:Overview_inner_faint}, strong magnetic field deviations are observed from a few \si{nT} to several hundreds of \si{nT}. These substantial deviations are mainly mapped to intense auroral lines in the \ac{UVS} images, as seen in Figure \ref{fig:Overview_inner_faint} (E), marked by the black vertical dashed lines. In a duration of \SI{2}{\min}, the magnetic field $\delta B_{\phi}$ increases by \SI{260}{nT} and then drops by \SI{300}{nT} as it approaches smaller L-shell values down to $12$. The passage of the spacecraft through strong changes in the magnetic field consistent with a curl of $\vec{B}$ (as discussed in Section \ref{subsec:static})  is indicative of a large-scale current region. Small-scale magnetic field fluctuations are hardly seen throughout this region at low altitudes. 

The positive slope indicates an upward current region with a characteristic downward filled electron loss cone and mono-energetic high electron intensities of \SI{100}{keV}, within an inverted V structure, commonly associated with a static electric potential. These structures are mapped to \ac{UVS} a bright auroral arc in the \ac{UVS} image. The negative slope at 04:42 in Figure \ref{fig:Overview_inner_faint} indicates a downward current region, mapping to dim \ac{UV} emissions, and demonstrates a filled upward loss cone but with a wide range of energy levels and high intensities, described as a broadband distribution. 

Although this is a compelling example, most unidirectional pitch angle distributions in the downward direction exhibit broadband energy distributions rather than monoenergetic structures. Unidirectional pitch angle distributions are commonly seen when the spacecraft crosses intense auroral regions, especially when they cover a small latitudinal range. The auroral crossings at greater radial distances also show mostly unidirectional magnetic field changes accompanied by intense auroral \ac{UV} emissions matching upward currents. Those magnetic field perturbations are smaller than those at low altitudes, usually observed with amplitudes ranging from a few tens of \si{nT} to no more than \SI{100}{nT}. 

% Absence of beams with large-scale deviations

Not every magnetic field variation exceeding an amplitude of \SI{50}{nT} leads to unidirectional electron pitch angle distributions or increased particle intensities. For example, during the first perijove crossing of the northern auroral zone, a notable magnetic field variation of \SI{-200}{nT} observed at 12:12 signifies an upward electric current, but the loss cones of the pitch angle distribution are vacant at that time, as shown in the supplementary Figure S3. Concurrently, the \ac{UVS} images exhibit intense auroral emissions. Hence, it is crucial to remember that the time discrepancy between instantaneous and time-averaged measurements, as well as the inaccuracy of the magnetic field-line mapping, might distort the comparison of different observations. Although unlikely, it is possible that the \ac{JEDI} might fail to detect electrons that have already moved to radial distances closer than the spacecraft's location or penetrated the atmosphere. Moreover, \ac{JEDI} might not detect electrons that are strongly aligned with the magnetic field within the resolved pitch angle distribution.

% Bi-direction and broadband electron distributions

\begin{figure}[ht!]
    \centering
    \includegraphics[width=\textwidth]{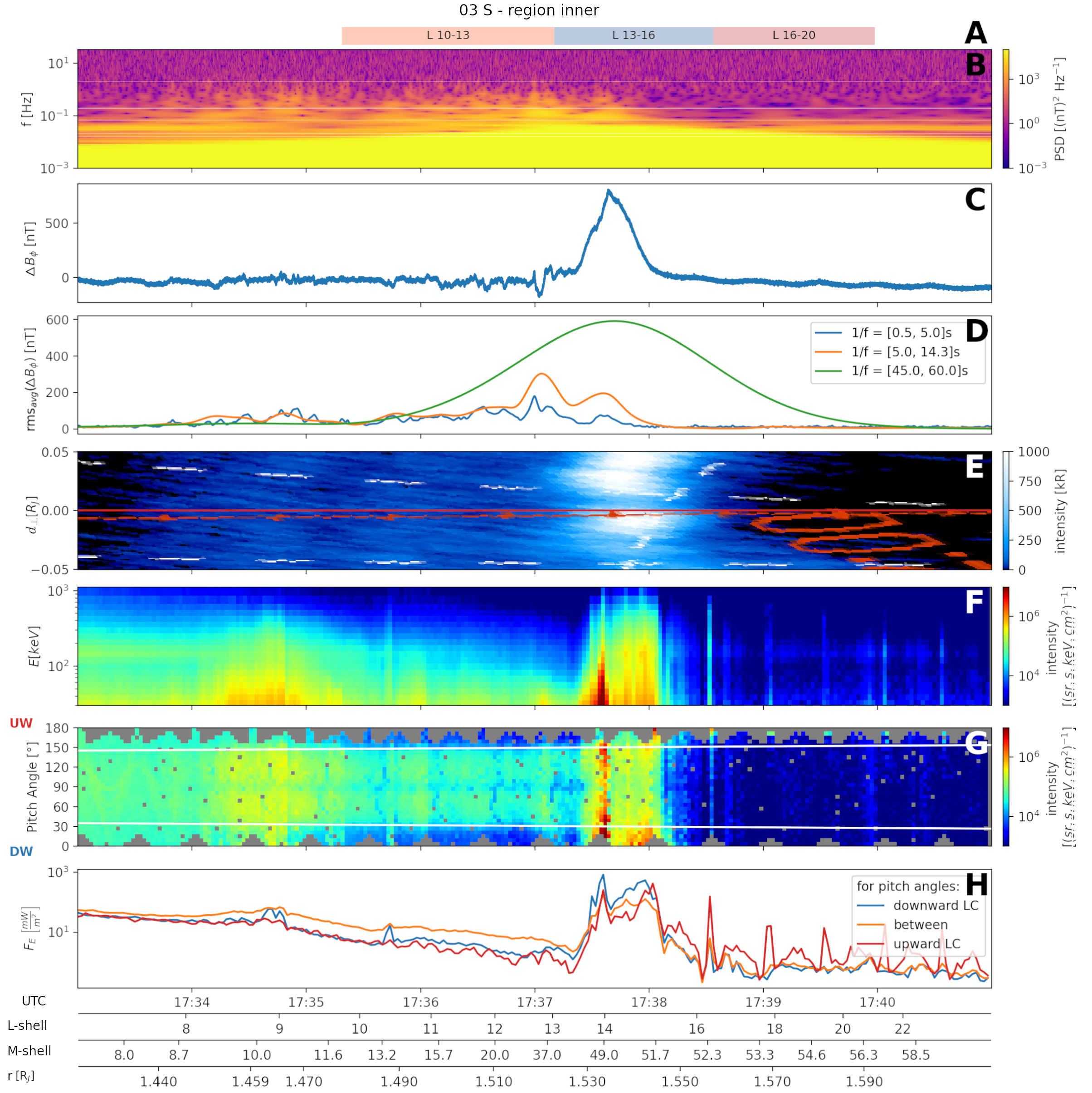}
    \caption[Multi instrument data for southern Perijove 3 at a radial distance \SI{<2}{R_J}]{This figure gives an overview of the instrument data from perijove 3 over the southern hemisphere passing at altitudes lower than \SI{1}{R_J}. Further details for each panel are similar to those provided in the caption of Figure \ref{fig:Overview_inner_faint}. 
    % The vertical dashed black lines indicate regions of intense aurora and correspond to strong magnetic field changes.
    }
    \label{fig:3S_low}
\end{figure}

Bidirectional pitch angle distributions are often observed in the smooth transition from \ac{ZI} to \ac{ZII}, as seen in the supplementary Figure S6 at 01:15. These regions are indicated by the red bar at the top of the Figures from L-shell \SIrange{16}{20}{}. These distributions exhibit a comparable energy flux in both cones, exceeding that outside the cones. Interestingly, bidirectional distributions often coincide with magnetic field changes indicating downward currents, alongside very intense auroral \ac{UV} emissions, as seen in Perijove 3 at 17:38 in Figure \ref{fig:3S_low}. This contradicts the idea that a static potential mainly accelerates the electrons since the electrons move in both directions independently of the possible current direction determined by the magnetic field change. The considerable amount of particles that are accelerated downward can cause the most intense auroral emission. Thus, bidirectional acceleration in an area connected to downward currents significantly contributes to the intense \ac{UV} emissions. 

These bidirectional distributions are also observable poleward through even higher L-shells. The \ac{UV} emissions show intensities that are bright and patchy with high intensities or diffuse with intermediate intensities. The magnetic field shows mainly negative changes of several \SI{100}{nT}, corresponding to a downward current. The electron distributions in these areas are similar in intensity to those of \ac{ZI}. However, the pitch angle distributions are usually bidirectional, sometimes almost isotropic, or unidirectional in an upward direction. At the same time during which the electron beams appear, negative magnetic field changes, accompanied by intense \ac{UV} radiation, are observed. As already observed in \ac{ZI}, if the pitch angle distributions are bidirectional, this \ac{UV} radiation is present (see supplementary Figure S5 at 06:58), but when the electrons only fill the upward loss cone, nearly no \ac{UV} emissions are visible (see supplementary Figure S4 at 09:47). 

The bidirectional distribution weakens in intensity towards the poles and sometimes transitions into distributions that are solely directed upward. However, electrons with energies up to \SI{1}{MeV} can reach the detectors at higher latitudes, as evidenced by the minimum ionizing effect observed at \SI{150}{keV}. Despite the bidirectional acceleration of the electrons, only faint \ac{UV} emissions are detected. Surprisingly, there are no observations of either small-scale magnetic field fluctuations or significant magnetic field changes. The acceleration in both directions could be attributed to small fluctuations that are too weak to be identified in the inner radial regions.

\subsection{Small-Scale Magnetic Field Fluctuations}

During all perijove observations, fluctuations in the magnetic field were detected on short time scales of a few seconds to several minutes and across various radial distances, particularly at L shells below $13$. These regions predominantly exhibited pancake distributions along with some diffuse auroral beams. The magnitude of these magnetic field fluctuations ranged from a few \si{nT} and tens of \si{nT} at higher altitudes (above \SI{1}{R_J}) to \SIrange{30}{600}{nT} at lower altitudes (below \SI{1}{R_J}). These high amplitude changes occur distinct from the Io-Alfvén wing crossings and are exemplarily shown in the supplementary Figure S1 between 9:36 and 9:39.

Most low-altitude crossings did not detect small-scale fluctuations in \ac{ZI} and \ac{ZII}. Nevertheless, particular powerful electron beams could not be correlated to any perturbation in the magnetic field, thus providing no direct insight into the acceleration mechanisms of these intense beams. In some cases, the magnetic field showed small-scale fluctuations of several seconds at low altitudes that directly contributed to intense emissions above the main emission zone, as seen in the \acf{UVS} observations, Figure S1 (B) between 17:26 and 17:27. 
Unfortunately, the electron distribution during this perijove was not fully captured, especially the upward loss cone. However, pitch angle distribution indicates diffuse emissions, with the upward loss cone empty and the downward loss cone and pitch angles outside the cone filled, as depicted in Figure S1 between 17:26 and 17:30. In particular, at 17:25, the electron energy distribution displays a beam with an inverted V structure, yet no magnetic changes indicative of field-aligned currents are detected. Additionally, no magnetic field fluctuations accompany downward or bidirectional electron beams, suggesting either insufficient resolution of magnetic field components to discern an acceleration mechanism or the spacecraft's position lying below the acceleration region.
Only six other perijoves show small-scale fluctuations for higher L shells that map to intense \ac{UV} emissions at these small radial distances. The amplitudes of the fluctuations are very small, almost equivalent to the high noise level of \SI{25}{nT} at these small radial distances. 

\ac{MAG} observations at small radial distances with less than \SI{2}{R_J} do not allow observation of fluctuations with less than \SI{10}{(nT)^2/Hz}, mostly corresponding to frequencies greater than \SI{0.8}{Hz}. However, beyond a distance of \SI{4}{R_J}, digitization by the \ac{FGM} results in noise levels below \SI{0.1}{nT} (as illustrated in Figure \ref{fig:correlation_dB_CWT} (B)), allowing the detection of smaller amplitudes associated with small-scale fluctuations.
A drawback of observations at larger radial distances is the changing temporal resolution of \ac{MAG} data, which sometimes decreases from \SI{64}{Hz} to \SI{16}{Hz}.

\begin{figure}[ht!]
    \centering
    \includegraphics[width=\textwidth]{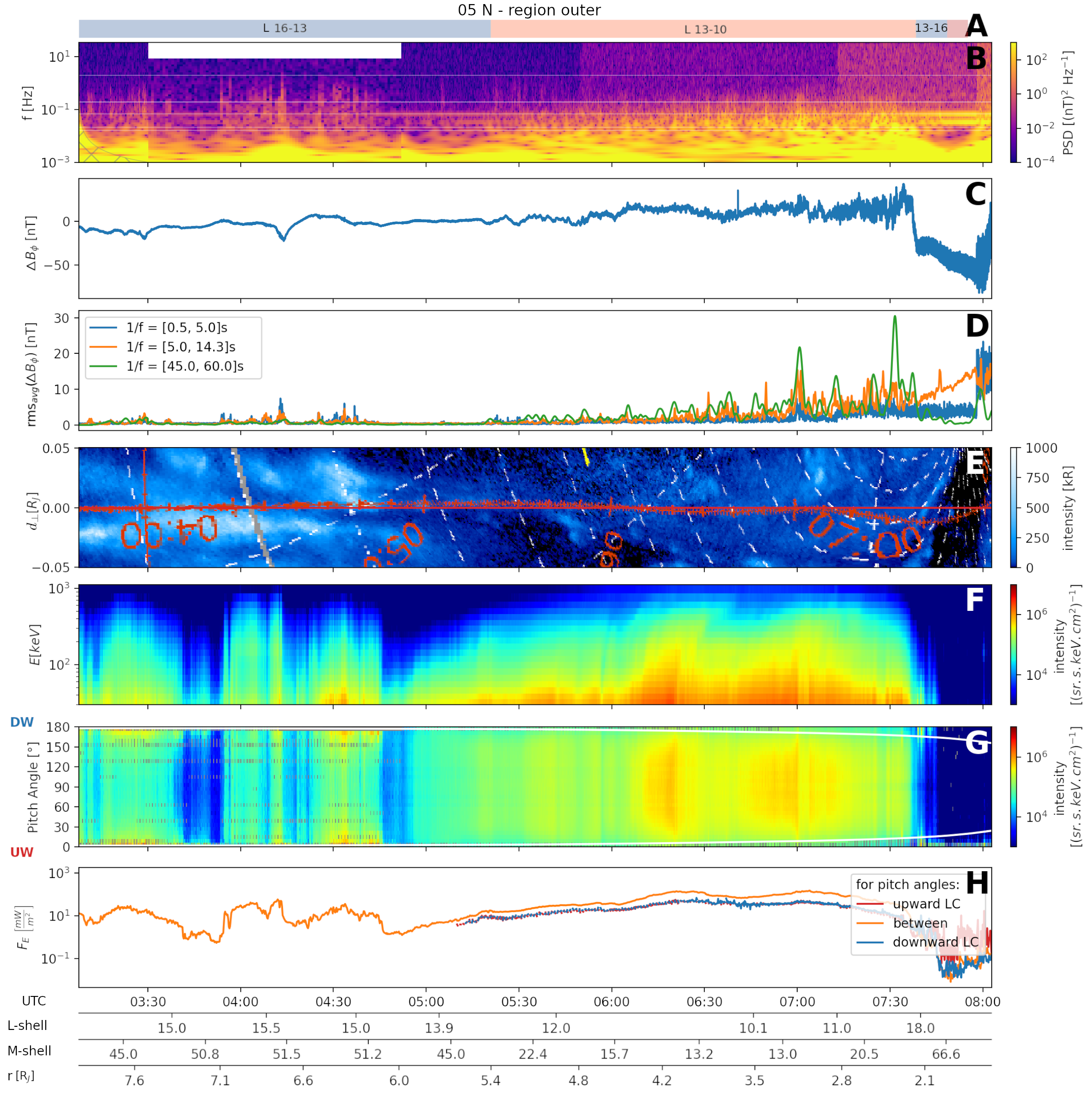}
    \caption[Multi-instrument data for northern Perijove 5 at a radial distance \SI{>2}{R_J}]{This figure gives an overview of the magnetic field, energetic electrons, and UV emission data from Perijove 5, which passed through the northern hemisphere at altitudes higher than \SI{1}{R_J}. Further details of each panel are provided in the caption of Figure \ref{fig:Overview_inner_faint}. }
    % The vertical dashed black lines indicate regions of intense aurora and correspond to strong magnetic field changes.
    \label{fig:Overview_fluc_outer-small1}
\end{figure}

Hence, small-scale fluctuations along higher radial distances \SI{>3}{R_J} typically show a rolling \ac{rms} of a few \si{nT} between \SIrange{0.5}{15}{\sec}. In particular, the fifth perijove, depicted in Figure \ref{fig:Overview_fluc_outer-small1} (B) (3:10-4:50) at radial distance of \SI{\sim 7}{R_J}, demonstrates small amplitude magnetic field fluctuations at temporal scales as short as \SI{0.5}{\second} when the noise level is low (\SI{0.024}{nT}). The \ac{rms} values of these fluctuations within \SIrange{0.5}{5}{\second} is about \SIrange{2}{5}{nT}, which are too small to be resolved at low altitudes due to the higher noise level. 

We estimate the corresponding energy flux at Jupiter’s ionosphere using \( F_E = \frac{(\delta B)^2}{\mu_0} \cdot c \cdot \frac{B}{B_m} \), where \( B \) is the local magnetic field and \( B_m \) is the traced ionospheric magnetic field. We approximate the Alfvén velocity with the speed of light, as expected for the densities observed in the previous paragrah \cite{Bagenal2014}. Based on the observed magnetic field fluctuations of approximately \SIrange{2}{5}{nT} at a radial distance of \SI{\sim 7}{R_J} (as described above), this yields a range of about \SIrange{380}{2370}{mW/m^2}. While these energy fluxes represent simplified estimates, it suggests that the observed magnetic field fluctuations can carry sufficient energy to drive structured auroral emissions—particularly since some fluctuations reach amplitudes of up to \SI{10}{nT} (see Supplementary Figure~S5 at 11:15).
\citeA{Lorch2022} report magnetic field fluctuations in the range of \SIrange{0.16}{0.675}{nT}, corresponding to projected ionospheric energy fluxes of approximately \SIrange{6.8}{196}{mW/m^2}, which are already sufficient to generate aurora. In our case, we estimate comparable or even significantly higher fluxes of several hundred \si{mW/m^2} up into the range of \si{mW/m^2}, indicating that the observed fluctuations are well within the energetic range required to support structured auroral processes, assuming effective transmission.
Similar small-scale fluctuations are observed shown in Supplementary Figure S2 between 10:30 and 11:15.

Broadly distributed electron populations at large radial distances  \SI{>3}{R_J} are associated with higher \ac{rms} values in the magnetic field fluctuations on the \SIrange{8}{20} {\second} scale. At high altitudes, distributions such as pancake, butterfly, and isotropic populations are difficult to distinguish from one another, as the loss cone is extremely small at these distances. However, the magnetic field fluctuations for these populations are large enough to remain visible despite the high noise levels.  This is shown in Figure \ref{fig:Overview_fluc_outer-small1} between 5:30-7:40.

We used wavelet analysis to examine the fluctuations in the magnetic field perpendicular to the magnetic field by categorizing their strength according to different time scales. The identified frequency bands strongly depend on the time that Juno spent within the main emission region, the constraints imposed by digitization, and the notable influence of the spacecraft's rotation. This influence becomes more pronounced with higher magnetic field strengths at smaller radial distances where digitization levels are already high, resulting in significant \aclp{PSD} of around \SI{1e3}{(nT)^2/Hz}.

\begin{figure}[ht!]
    \centering
    \includegraphics[width=\textwidth]{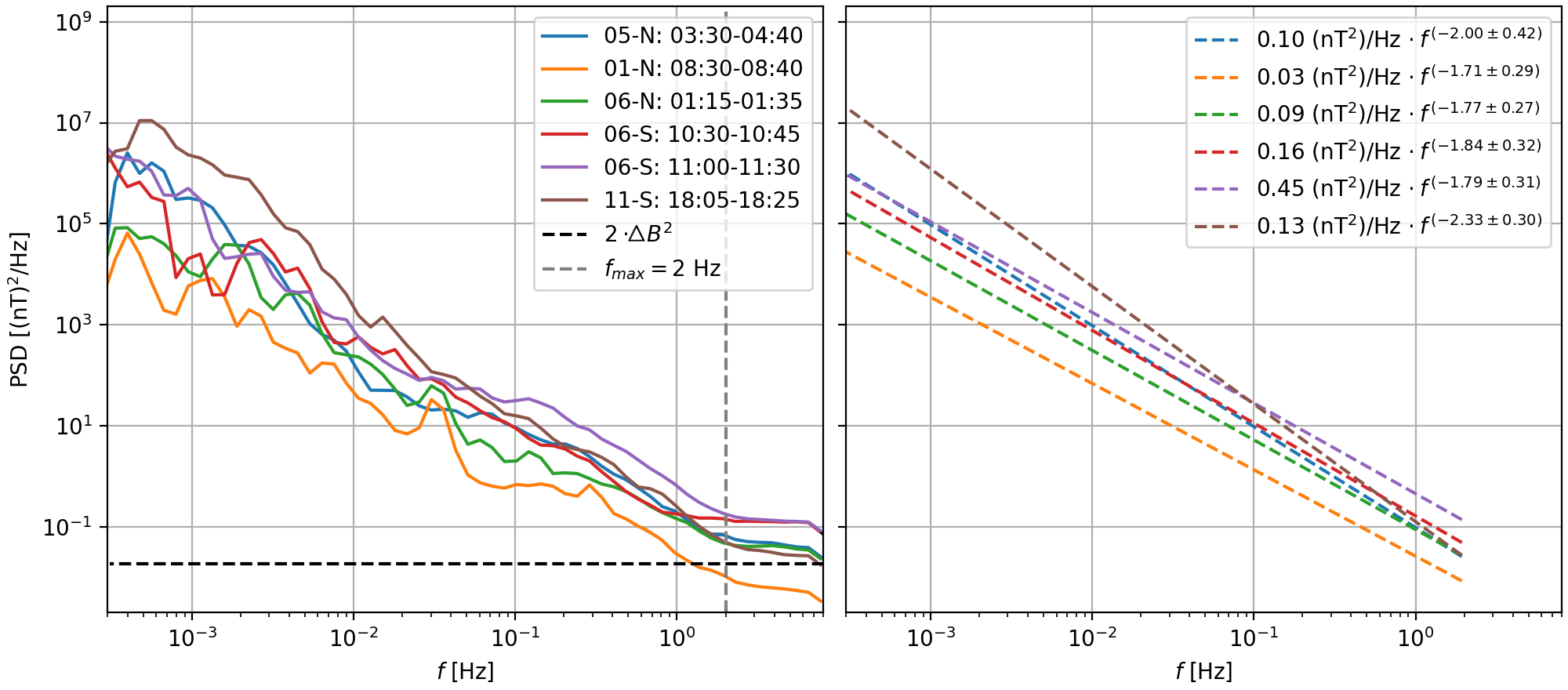}
    \caption[Power spectra for six main emission crossings at high altitudes]{Six distinct time spans at high radial between \SIrange{6}{7.5}{R_J} distances crossing the main emission zone. The power spectra of magnetic field fluctuations measured during these time periods are shown, each providing evidence for small-scale fluctuations up to the highest frequencies. The right-hand side displays the linear regression of the time-averaged spectra on the left. The black and grey dashed line indicate the boundary to unresolved frequencies for the corresponding digitization levels.}
    \label{fig:small}
\end{figure}

Measurements obtained with a digitization level of \SI{0.1}{nT} are possible at radial distances greater than \SI{5}{R_J}. The spacecraft rarely crosses the main emission region at these altitudes. Nevertheless, six distinct time intervals for observing intense ultraviolet emissions are available, as marked by the black lines within the orange zone of Figure \ref{fig:data}. Figure \ref{fig:small} illustrates the power spectral densities measured over these periods, which exhibit a power-law like distribution with fluctuations reaching into high-frequency ranges. The corresponding slopes, shown on the right, range between $1.71$ and $2.33$, with a power spectral density (\acl{PSD}) between \SIrange{0.03}{0.45}{(nT)^2/Hz} at a frequency of \SI{1}{Hz}. The gray and black dashed lines in Figure \ref{fig:small} show the digitization level and the corresponding boundary of the resolved frequencies. Certain time periods exhibit significant magnetic field fluctuations at lower frequencies with amplitudes reaching up to \SI{50}{nT}. As a result, the corresponding spectra are steeper, offering greater power at frequencies below \SI{0.1}{Hz}.

All of the six time intervals at larger distances have bidirectional electron distributions or isotropic distributions with unresolved loss cones. The energy distribution is broad and the ultraviolet emission is intense, reaching up to \SI{1000}{kR}. The energy fluxes are mostly in the range of a few tens of \si{mW/m^2} but can reach a few hundred \si{mW/m^2} when amplitudes of the magnetic field fluctuations are greater than \SI{10}{nT}. 

These perijoves display small-scale fluctuations at the lowest digitization levels at a radial distance between approximately \SIrange{6}{7.5}{R_J} and L-shell between $12.5-17$, each time the digitization level allows to resolve them. This suggests that small-scale fluctuations might be an omnipresent feature of Jupiter's magnetosphere-ionosphere coupling. This would further stress the importance of stochastic acceleration, consistent with the wide range of auroral particle observations at low altitudes.

% \begin{figure}
%     \centering
%     \includegraphics[width=\textwidth]{images/1.png}
%     \caption{mean rms through all 20 perijoves}
%     \label{fig:rms_overview}
% \end{figure}

%%%%%%%%%%%%%%%%%%%%%%%%%%%%%%%%%%%%%%%%%%%%%%%
\section{Summary and Conclusions}\label{sec:summary}

In this study, a comprehensive analysis was conducted on the Juno observations during the first 20 flybys where magnetic fields, electron intensities, and \ac{UV} emission were systematically compared. Our findings confirm the characterization by \citeA{Mauk2020} of the auroral emission region in the three zones but provide an additional new and significant understanding of \ac{MI}-coupling processes associated with magnetic field fluctuations.

\textbf{Polewards of the main emission}, predominantly upward and bidirectional electron beams are detected, with occasional faint ultraviolet emissions. No magnetic field fluctuations were detected at radial distances between \SIrange{1.25}{3}{R_J}, suggesting that any fluctuations might be too small to observe, though they may contribute to acceleration via whistler-mode waves, as noted by \citeA{Elliott2020}.

However, luminous emissions are also observed in the polar region area, where three distinct bright auroral spots have been examined by \citeA{Haewsantati2023}.
Juno's observations traced along the magnetic field lines to these spots revealed an enhanced upward flow of energetic electrons, increased whistler-mode wave activity, and fluctuations in the magnetic field. The results suggest that beneath Juno, particle acceleration occurs, and wave-particle interactions play a role in creating UV aurora, consistent with observations by \citeA{Mauk2020}. 
This is consistent with \citeA{Gerard2019}, who observed an inconsistency between electron flux and UV brightness.

\textbf{The diffuse auroral emission zone} is characterized by pancake-shaped electron distributions. In the diffuse emission zone, field lines demonstrate minimal significant fluctuations in the magnetic field perpendicular to the background field, when considering the noise level. Electron intensities sometimes fill the downward loss cone, along with magnetic field fluctuations that are perpendicular to the background field and occur at frequencies between $1/60$ and \SI{0.5}{Hz}. The amplitudes of these fluctuations show a radial dependence, reaching several hundred nanoteslas (\si{nT}) at radial distances less than \SI{4}{R_J} and tens of \si{nT} beyond \SI{4}{R_J}.

\textbf{The main emission zone} shows large-scale magnetic field variations correlated with unidirectional and bidirectional electron distributions, as defined in \citeA{Mauk2020}. The changes in the magnetic field perpendicular to the background magnetic field occur on large scales corresponding to tens of minutes with magnetic field amplitudes of several \SI{100}{nT} and are consistent with quasi-static large-scale electric currents. Consequently, the downward electron distributions align with the upward current regions, which are linked to strong \ac{UV} emissions. In contrast, areas with downward currents sometimes show only upward-moving electrons without \ac{UV} emissions. However, most of the time regions with downward currents exhibit electrons moving in both directions, leading to significant \ac{UV} emissions.
Therefore, the prevalence of broadband distributions and the dominance of bidirectional electron flow in these regions seem inconsistent with the explanation of static potential drops \cite{Salveter2022}. 

Fluctuations in the magnetic field on smaller temporal scales manifest exclusively in areas linked to diffuse auroral regions at lower altitudes and diminish upon entering regions connected to the main emission zone \cite{Sulaiman2022}. \citeA{Sulaiman2020} has proposed that this lack of fluctuations could be due to significant reductions in density above the main emission zone, resulting in large electron inertial lengths scales and thus more efficient wave-particle interaction \cite{Saur2018, Lysak2021}. On the other hand, the absence of magnetic field fluctuations might be attributed to the inability to detect minor fluctuation amplitudes. This occurs as a result of the reduction in density when considering the same Poynting flux $\delta B^2/\mu_0 \cdot v_A$ and leading to significant increases in the Alfvén speed and therefore decrease in $\delta B$.

An extensive analysis of the power spectrum has revealed that the transition from analog to digital measurement (i.e. digitization) significantly restricts the observable temporal scales of the magnetic field data. Specifically, when observing radial distances greater than \SI{4}{R_J} as opposed to radial distances less than \SI{2}{R_J}, the magnetic field strength decreases by approximately two magnitudes. As a result, there is an improvement in the resolution of the magnetic field measurements from $25$~nT to $0.1$~nT. Thus, the digitization level with the highest resolution (that is, the lowest digitization level) is capable of detecting frequency changes above \SI{0.5}{Hz}. 
This level of resolution is only used at radial distances larger than \SI{4}{R_J}. Consequently, the identification of small magnetic field fluctuations can occur only at greater radial distances. 

Throughout the mission, the trajectory of the Juno spacecraft evolves due to the gravitational influence of Jupiter’s oblateness, which causes Juno's perijove latitude to shift northward by approximately $1$ degree per orbit, starting with a near-equatorial perijove. This results in relatively brief observation periods connected to the main emission at higher altitudes in the north and extended periods connected to high latitudes during Juno’s outward passages.

However, the perpendicular component of the magnetic field consistently exhibits small-scale fluctuations at radial distances larger than \SI{4}{R_J}, when mapping to intense \ac{UV} emissions in the main emission zone.  
The slope of the \ac{PSD} resembles a power law with a slope ranging from $-1.7$ to $-2.2$, extending to higher frequencies of up to \SI{2}{Hz}, which is consistent with the power spectral densities discovered by \citeA{Lorch2022} at radial distances greater than \SI{10}{R_J}. A power law in the spectral energy density is indicative of turbulent fluctuations \cite{Saur2002, Tao2011}. Amplitudes of \SI{2}{nT} to \SI{10}{nT} of these fluctuations carry very large energy fluxes on the order of \SI{0.1}{W/m^2} to several \si{W/m^2} when mapped to Jupiter.

%%%%%%%%%%%%%%%%%%%
%What you found: Results – What will you conclude? What is the point of your paper?
% Minor magnetic field fluctuations, observed even at short intervals, imply additional auroral electron acceleration, likely due to wave interactions.
We conclude that electrostatic field-aligned currents, as well as wave-particle interaction, may contribute significantly to intense auroral arcs on Jupiter. A stochastic acceleration process seems to drive the dominant amount of particle observation \cite{Saur2018, Salveter2022, Lorch2022, Gershman2019, Damiano2019, Damiano2023}. The coexistence of these accelerations underscores Jupiter's magnetospheric variability. Enhancing this analysis in spatial and temporal coverage using the Juno data and using the full range of Juno instruments would be valuable. 

Our study investigates the occurrence of large-scale magnetic field variations associated with quasi-static field aligned Direct Currents (DC) and small-scale magnetic field fluctuations associated with highly time-variable Alternating Currents (AC). We find in most cases large-scale current systems over the main auroral emission are present but also detect that alternating currents are always present when they can be measured. The latter observations in conjunction with the fact that the dominant electron distribution functions are broadband over the main emission zone suggestive of wave-particle interaction being the dominant acceleration mechanism \cite{Saur2018, Salveter2022, Lorch2022, Gershman2019, Damiano2019, Damiano2023, Lysak2021}. Further theoretical and observational studies based on a large suite of Juno instruments and extending it to further perijoves will help to bring further light into understanding Jupiter's enigmatic auroral emissions.

%%%%%%%%%%%%%%%%%%%%%%%%%%%%%%%%%%%%%%%%%%%%%%%
\section{Open Research}
 The Juno-\ac{JEDI} data  \cite{JEDIdata} as well as the Juno-Magnetometer data \cite{System2022} were obtained from the website of the NASA Planetary Data System:Planetary Plasma Interactions (\url{https://pds-ppi.igpp.ucla.edu/mission/JUNO/JNO/JEDI} and \url{https://pds-ppi.igpp.ucla.edu/mission/JUNO/JNO/FGM}). The Juno-\ac{UVS} calibrated data \cite{UVSdata, UVdata} can be utilized from \url{https://pds-atmospheres.nmsu.edu/cgi-bin/getdir.pl?dir=DATA%26volume=jnouvs_3001} to obtain the polar projection images described by \citeA{Bonfond2021}. 
 Juno footprints are available using \ac{JRM09} and \ac{Con2020} models at \url{https://lasp.colorado.edu/home/mop/files/2020/04/20190412_Imai_MagFootReader_UIowa_rev.pdf}. The tracking of the magnetic field lines with \ac{JRM33} and \ac{Con2020} is provided by \cite{Wilson2023}. The classification results for calculating the precipitation budget of the different electron distributions are provided as supplementary material in \citeA{Salveter2022}.

%%%%%%%%%%%%%%%%%%%%%%%%%%%%%%%%%%%%%%%%%%%%%%%
\acknowledgments
Annika Salveter and Joachim Saur acknowledge the funding from the Deutsche Forschungs Gemeinschaft for their support through the special program Jupiter's aurora: Data analysis of Juno/JEDI data and modeling of auroral electron acceleration (SA 1772/6-1). 

%%%%%%%%%%%%%%%%%%%%%%%%%%%%%%%%%%%%%%%%%%%%%%%
\bibliography{references}

\end{document}